\documentclass[letter]{ptptex}
\usepackage{wrapft}
\usepackage{graphicx}

\title{
Positive Energy Theorem Implies Constraints on Static Steller Models 
}

\author{
Tetsuya \textsc{Shiromizu}$^{1}$
and Hirotaka \textsc{Yoshino}$^{2}$
}

\inst{
$^1$Department of Physics, Tokyo Institute of Technology, Tokyo 152-8551, Japan\\
$^2$Graduate School of Science and Engineering, Waseda University, Tokyo 169-8555, Japan}

\abst{
Using the positive energy theorem,
we derive some constraints on static steller models in asymptotically flat spacetimes 
in a general setting without imposing spherical symmetry. 
We show that there exist no regular solutions under certain 
conditions on the equation of state. As the contraposition, we obtain some 
constraints on the pressure and adiabatic index.  
}

\begin{document}

\maketitle


1. {\it Introduction}. In this Letter we consider certain constraints on 
general relativistic 
static steller models that are implied by the positive energy theorem \cite{SY},
without assuming spherical symmetry. 
Although it has been proven that a static star is spherically 
symmetric under some reasonable assumptions \cite{Sph}, it is still interesting 
to obtain the constraints without imposing any symmetries. 
This kind of 
argument was first considered by Lindblom and Masood-ul-Alam \cite{LM} (See also Ref. \cite{Simon}). 
In this 
Letter we reconsider their argument and derive significantly stricter 
and different constraints on the pressure and the adiabatic index.

The metric of a static spacetime is given by 
%
\begin{eqnarray}
ds^2=-V^2(x)dt^2+g_{ij}(x)dx^i dx^j.
\label{metric}
\end{eqnarray}
%
We consider steller models composed of a barotropic perfect fluid with 
energy density $\rho\ge 0$ and pressure $P=P(\rho) \geq 0$. 
The Einstein equations and fluid equation for such a system are given by 
%
\begin{eqnarray}
& & D^2 V =4\pi V (\rho + 3P), \label{00}  \\ 
& & R_{ij}=\frac{1}{V}D_i D_j V + 4\pi (\rho -P)g_{ij}, \label{ij}\\ 
& & D_i P = -\frac{1}{V}(\rho +P) D_i V, \label{pf} 
\end{eqnarray} 
%
where $D_i$ and $R_{ij}$ are the covariant derivative and 
Ricci tensor of the metric $g_{ij}(x)$. 
Since we do not consider the existence of the 
event horizon, 
$V$ is larger than zero and smaller than unity. 
Equation \eqref{pf} shows that $P$ and $\rho$ are regarded as 
functions of $V$ 
and the star surface corresponds to $V=V_s$. We denote 
the minimum value of $V$ inside the star as $V_c$. Then 
$V_s$ and $V_c$ satisfy the relation $0<V_c<V_s<1$. 

Asymptotic flatness requires that $V$ asymptotically behave as   
%
\begin{eqnarray}
V = \left( 1-\frac{M}{r} \right) +O(r^{-2})
\end{eqnarray}
%
and the metric be given by 
%
\begin{eqnarray}
g_{ij} = \left( 1+\frac{2M}{r} \right) \delta_{ij} +O(r^{-2}),
\end{eqnarray}
%
where $r=:|\delta_{ij}x^i x^j|^{1/2}$ and $M$ is the ADM mass. 
Because $R=16\pi \rho \geq 0$, 
the positive energy theorem shows the positivity of $M$. 

In the next two sections, we consider the conformal transformation
%
\begin{eqnarray}
\tilde g_{ij}=\omega^{4}g_{ij}, 
\label{conformal-transformation}
\end{eqnarray}
%
where we assume that $\omega$ is a function of $V$: $\omega=\omega(V)$. 
The Ricci scalar is transformed as 
%
\begin{eqnarray}
\tilde R  =  \frac{1}{\omega^{4}}
\Bigl[
R-8\frac{D^2 \omega}{\omega}
\Bigr] = 
\frac{1}{\omega^{4}}
\Biggl\lbrace
16\pi
\left[
\rho-2 \frac{\omega_{,V}}{\omega}V(\rho+3P)
\right]-
\frac{8\omega_{,VV}}{\omega}
(DV)^2
\Bigg\rbrace,
 \label{ricci}
\end{eqnarray}
%
where $\omega_{,V}=d \omega /dV$ and $\omega_{,VV}=d^2 \omega /dV^2$. 
We require $\omega(1)=1$, and thus
the asymptotic behavior of the conformally transformed metric becomes 
%
\begin{eqnarray}
\tilde g_{ij} = \left(1+\frac{2 \tilde M}{r}  \right) \delta_{ij} +O(1/r^2),
\end{eqnarray}
%
where the ADM mass $\tilde{M}$ 
of the conformally transformed 
spacetime is given by 
%
\begin{eqnarray}
\tilde M=M(1-2\omega_{,V}(1)). 
\end{eqnarray}
%
We note that the conformally transformed spacetime has a 
non-positive ADM mass if 
%
$\omega_{,V}(1) \ge  \frac12. $
%


2. {\it Constraints on the pressure}. 
Let us begin by deriving constraints on the pressure. 
This problem is easier than in the case of the adiabatic index, 
and it is useful to understand the strategy for obtaining constraints 
on that using the positive energy theorem.   
We choose $\omega$ such that $\omega_{,V}(1)\ge 1/2$ 
and $\omega_{,VV}\le 0$ for $V_c\le V\le 1$. Then, if the inequality 
%
\begin{eqnarray}
\frac{P}{\rho}<\frac13\left(\frac{\omega}{2V\omega_{,V}}-1\right) 
\label{pressure-cond}
\end{eqnarray}
%
is satisfied inside the star, we have $\tilde R \geq 0$ everywhere.

Now we can use the positive energy theorem. Because 
the ADM mass of the conformally transformed spacetime is non-positive, 
it turns out 
that the conformally transformed space is flat. Thus we have $\tilde{R}=0$, 
and Eq.~\eqref{ricci} implies that $V$ is constant. 
Hence, Eq. (\ref{00}) indicates $\rho=P=0$, 
and Eq. (\ref{ij}) implies that $R_{ij}=0$. Because the space is 
three dimensional, we have $R_{ijkk}=0$. 
Thus, the original space is also Euclid space. 
As a consequence, we find that there is no solution of the 
static steller models satisfying the inequality \eqref{pressure-cond}. 
In other words, Eq. \eqref{pressure-cond}
should be violated at some point for any solution 
of static steller models.

We consider the choice of $\omega$ that leads to the optimal condition.
From Eq. \eqref{pressure-cond}, it is expected that the optimal condition is 
realized when $\omega$ is maximal 
and $\omega_{,V}$ is minimal. 
Clearly, this corresponds to the choice
%
\begin{equation}
\omega=\frac12(1+V),
\label{omega-choice-optimal}
\end{equation}
%
because if we choose some other $\omega$ such that $\omega_{,VV}<0$,
the value of $\omega$ becomes smaller than $(1/2)(1+V)$, 
and $\omega_{,V}$ becomes larger than $1/2$.

Adopting Eq. \eqref{omega-choice-optimal},
we obtain the following. \\ 
{\it Theorem: Consider the static steller models
given by the metric \eqref{metric} in the asymptotically-flat spacetimes.
There is no solution satisfying 
%
\begin{equation}
\frac{P}{\rho} < \frac{1-V}{6V}.
\label{pressure-constraint-optimal} 
%
\end{equation} 
}
As the contraposition to this theorem, we also obtain the following:
{\it The solution of the static steller models given by the metric \eqref{metric}
should violate the above inequality at some point. } 

Strictly speaking, for the choice \eqref{omega-choice-optimal},
$\tilde R=0$ does not necessarily imply $D_i V=0$ or $\rho=P=0$, 
because we have $\omega_{,VV}=0$. 
However, the constraint \eqref{pressure-constraint-optimal} is actually derived as a limit of
$\omega$ that satisfies $\omega_{,VV}<0$. For example,
one can choose $\omega=-a(V-1)^2+\frac12(V+1)$ with $a>0$
and substitute this into \eqref{pressure-cond}. Then, taking the limit $a\to 0$,
the constraint \eqref{pressure-constraint-optimal} is found.

Our result indicates that the gravitational potential $V$ 
should not be very ``deep'' in the case that the pressure $P$ 
cannot take large values; i.e., the equation of state is soft.


3. {\it Constraint on adiabatic index}. In the previous section, in order to keep the condition 
$\tilde R \geq 0$, 
we introduced  
$\omega$ for which the relation $\omega_{,VV}\le 0$ is guaranteed
and derived some constraints by imposing the inequality 
$\rho>2(\omega_{,V}/\omega)V(\rho+3P)$.
In this section, we take opposite approach: We introduce $\omega$
for which the relation $\rho\ge 2(\omega_{,V}/\omega)V(\rho+3P)$ is guaranteed
and derive some constraints by imposing the inequality $\omega_{,VV}<0$. 
We find that this procedure yields constraints on the adiabatic index 
%
\begin{eqnarray}
\gamma= \frac{\rho+P}{P} \frac{dP}{d \rho}. 
\end{eqnarray}
%

We choose the conformal factor 
$\omega=(1/2)(1+V)$ outside the star. Inside the star, we choose 
%
\begin{eqnarray}
\omega = \frac12(1+V_s) {\exp} \left( \int^{V}_{V_s} 
\frac{f(x)}{V^n}dV \right), \label{omega}
\end{eqnarray}
%
where $f(x)$ is the following function of $x:=P/\rho$:
\begin{equation}
f(x):=\left(\frac{V_s^n}{1+V_s}\right)\frac{1}{1+3x}. \label{f}
\end{equation}
In Eqs. (\ref{omega}) and (\ref{f}), $n$ is a constant satisfying
%
\begin{equation}
1\le n\le n_{\rm max}:=1+\frac{\log\frac{1+V_s}{2V_s}}{\log\frac{V_s}{V_c}}.
\label{n-condition}
\end{equation}
%
The quantity $\omega_{,V}=\omega f(x)/V^n$ 
is continuous at $V=V_s$, because $P\sim \rho^\gamma$
and $x\sim \rho^{\gamma-1}\to 0$ at the star surface.
For this $\omega$, we obtain 
%
\begin{equation}
\rho-2\frac{\omega_{,V}}{\omega}V(\rho+3P)=
2\rho\left[\frac{1}{2}-\frac{V_s}{1+V_s}\left(\frac{V_s}{V}\right)^{n-1}\right]\ge 0,
\end{equation}
%
where we have used the relation $V_c\le V$ and the condition \eqref{n-condition} for $n$.
Then, calculating $\omega_{,VV}$, we find 
%
\begin{equation}
{V^{n+1}}\frac{\omega_{,VV}}{\omega}
=\frac{f^2}{V^{n-1}}-nf-(1+x)
\left(1-\frac{1+x}{\gamma}\right)f_{,x}.
\end{equation}
%
Thus, if $\gamma$ satisfies the inequality
%
\begin{eqnarray} 
\gamma  < \frac{-(1+x)^2f_{,x}}{V_c^{1-n}f^2-nf-(1+x)f_{,x}} =\frac{3(1+x)^2}{F(n)+3(1-n)x}
,\label{condition-gamma}
\end{eqnarray}
%
the relation $\omega_{,VV} <0$  holds, where $F(n)$ is defined by 
%
\begin{eqnarray} 
F(n):=\frac{V_s}{1+V_s}\left(\frac{V_s}{V_c}\right)^{n-1}-n+3.
\label{function-def}
\end{eqnarray}
%
In the above, we have again used $V_c\le V$ and Eq. \eqref{n-condition}.
From the above, it is seen that Eq. \eqref{ricci} 
implies $\tilde R \geq 0$ and $\tilde M=0$. Then, employing 
the same argument as in the previous section,
we find that the original space is a vacuum space, 
and thus static steller models satisfying the inequality~\eqref{condition-gamma}
do not exist, or equivalently, Eq. \eqref{condition-gamma} 
should be violated 
at some point if the solutions do exist.

Lindblom and Masood-ul-Alam proposed the constraint 
%
\begin{equation}
\gamma <\frac{3(1+V_s)}{2+3V_s}(1+x)^2,
\label{condition-LM}
\end{equation}
%
which corresponds to the $n=1$ case in 
Eq. \eqref{condition-gamma} \cite{LM}. 
By taking account of cases for which $n >1$, it is possible 
to derive a stricter constraint on $\gamma$. 
For this purpose, we choose the value of $n$ 
that minimizes the function $F(n)$ 
in order to make the right-hand side of Eq. \eqref{condition-gamma}
as large as possible. The derivative of $F(n)$ is given by 
%
\begin{equation}
F_{,n}(n)=\frac{V_s}{1+V_s}\left(\frac{V_s}{V_c}\right)^{n-1}\log\frac{V_s}{V_c}-1.
\end{equation}
%
In the case $V_c\le {V_s}/\exp({\frac{1+V_s}{V_s}})$, 
$F(n)$ takes its minimum value at $n=1$ in the range \eqref{n-condition}, because 
we have $F_{,n}(n_{\rm max})>F_{,n}(1)>0$.
In this case, the constraint we obtained on $\gamma$ is 
the same as that derived by Lindblom and Masood-ul-Alam, 
\eqref{condition-LM}. 

Next, we consider the case ${V_s}/\exp({\frac{1+V_s}{V_s}})\le V_c\le V_s/\exp(2)$.
In this case, we have $F_{,n}(n_{\rm max})>0>F_{,n}(1)$.
Thus, $F(n)$ takes its minimum value at a value of $n$ 
between $1$ and $n_{\rm max}$.
This minimum is located at 
%
\begin{equation}
n_*=1-\frac{\log\left(\frac{V_s}{1+V_s}\log\frac{V_s}{V_c}\right)}{\log\frac{V_s}{V_c}},
\end{equation}
%
and for this value of $n$, we have 
%
\begin{equation}
F(n_*)=2+\frac{1+\log\left(\frac{V_s}{1+V_s}\log\frac{V_s}{V_c}\right)}{\log\frac{V_s}{V_c}}.
\end{equation}
%

Finally, we study the case $V_s/\exp(2)\le V_c$. In this case, the
minimum value of $F(n)$ is located at $n=n_{\rm max}$, 
because $0>F_{,n}(n_{\rm max})>F_{,n}(1)$.
The minimum value of $F(n)$ is
%
\begin{equation}
F(n_{\rm max})=\frac52-\frac{\log\frac{1+V_s}{2V_s}}{\log\frac{V_s}{V_c}}.
\end{equation}
%
 
We can summarise the above results in the following \\
{\it Theorem: Let us consider the 
static steller models
in the asymptotically flat and static spacetime given by the metric ~\eqref{metric}.
Let $V_c$ be the minimum value of $V$ and $V_s$ be 
the value of $V$ at the surface of the star. 
Then there exists no solutions with the adiabatic index satisfying 
%
\begin{equation}
\gamma <\begin{cases}
\displaystyle
\frac{3(1+V_s)}{2+3V_s}(1+x)^2, 
&
\displaystyle
 \Bigl(V_c\le \frac{V_s}{\exp{\frac{1+V_s}{V_s}}}\Bigr),
 \\
\displaystyle
\frac{3\log\frac{V_s}{V_c}(1+x)^2}{2\log\frac{V_s}{V_c}+1
+\log\left(\frac{V_s}{1+V_s}\log\frac{V_s}{V_c}\right)(1+3x)}
&
\displaystyle
\Bigl(\frac{V_s}{\exp\frac{1+V_s}{V_s}}\le V_c\le \frac{V_s}{\exp(2)}\Bigr)
\\
\displaystyle
\frac{6\log\frac{V_s}{V_c}(1+x)^2}{5\log\frac{V_s}{V_c}-2\log\frac{1+V_s}{2V_s}
(1+3x)}.
&
\displaystyle
\Bigl(\frac{V_s}{\exp(2)}\le V_c\Bigr)
\end{cases}
\label{derived-condition}
\end{equation}
%
}
As the contraposition to this theorem, we have the following:
{\it In the asymptotically flat and static spacetime, 
the static steller models given by the metric \eqref{metric}
 violate the inequality \eqref{derived-condition}
at some point if the solutions exist.} 

We now comment some brief remarks. 
In the case $V_s/\exp(2)\le V_c$, 
the minimum value of $F(n_{\rm max})$ is negative 
for $V_c>V_s\left(\frac{2V_s}{1+V_s}\right)^{2/5}$.
Thus, $F(n)$ becomes zero
for some value of $n$ satisfying $1<n<n_{\rm max}$. 
Then, the contraposition to the theorem 
asserts that $\gamma$ must be larger than infinity at some point,
which is, of course, impossible.
Thus, we find that there may exist a solution of static steller models 
only in the case 
%
\begin{equation}
V_c<V_s\left(\frac{2V_s}{1+V_s}\right)^{2/5}. 
\label{allowed-VsVc}
\end{equation}
%
This condition is somewhat stricter than the simple condition $V_c<V_s$.
Applying the same argument
for the case ${V_s}/\exp({\frac{1+V_s}{V_s}})\le V_c\le V_s/\exp(2)$,
we find that solutions may exist only in the case 
\begin{equation}
\log\frac{V_s}{V_c}\left(\frac{V_s}{V_c}\right)^2
> \frac{1+V_s}{\exp(1)V_s}
\label{allowed-VsVc-2}
\end{equation}
from the requirement $F(n_*)>0$. 
Because the above inequality is
satisfied for $V_s>1/(2e^5-1)\simeq 0.0034$ 
in the region ${V_s}/\exp({\frac{1+V_s}{V_s}})\le V_c\le V_s/\exp(2)$, 
the condition \eqref{allowed-VsVc-2} excludes only a tiny portion
of this region.


Next we consider the Newtonian limit.
In this limit, the relation $0<1-V_s<1-V_c\ll 1$ and $x\ll 1$ hold. 
Then, the contraposition to the theorem asserts 
that $\gamma$ must satisfy
%
\begin{equation}
\gamma \ge \frac{6\frac{1-V_c}{1-V_s}-6}{5\frac{1-V_c}{1-V_s}-6}
\end{equation}
%
at some point. From this inequality, we can conclude at 
least the following three statements:
(i) the static solution is allowed only for 
$\frac{1-V_c}{1-V_s}\ge 6/5$; (ii) for arbitrary allowed values of 
$V_c$ and $V_s$, $\gamma$ must become larger than $6/5$ at some point; 
(iii) for the solution of the constant adiabatic index $\gamma=\gamma_*$, 
the relation 
\begin{equation}
\frac{1-V_c}{1-V_s}\ge\frac{6\gamma_*-6}{5\gamma_*-6}
\label{inequality-VcVs}
\end{equation}
must hold. 
Because a Newtonian static star is stable only for $\gamma>4/3$ \cite{Chandra},
the right hand side of the inequality \eqref{inequality-VcVs} 
takes a value between 6/5 and 3 for stable stars.

The derived universal condition $\gamma> 6/5$ is quite close to the
Newtonian result for the stability condition, $\gamma>4/3$. We 
believe that these two analyses are closely related, 
because the proof of 
the present theorem is based on the positive energy theorem, which has a 
close connection to the stability of spacetimes. However, 
elucidating this connection is beyond the scope of this Letter, and 
a direct approach 
based on perturbative analysis is needed.

\section*{Acknowledgements}

The work of T.S. was supported by Grants-in-Aid for Scientific 
Research from the Ministry of Education, Culture, Sports, Science and 
Technolohy of 
Japan(Nos.13135208, 14102004, 17740136 and 17340075) and  
the Japan-U.K. and Japan-France Research  Cooperative Programs. 
The work of H.Y. was partially supported by a Grant for The 21st Century 
COE Program (Holistic Research and Education Center for Physics 
and Self-Organization Systems) at Waseda University.


\begin{thebibliography}{99}

\bibitem{SY}
R. Schoen and S.-T. Yau, Commun. Math. Phys. {\bf 65}(1979), 45; ibid, {\bf 79}(1981), 231. 

\bibitem{Sph}
L. Lindblom and A. K. M. Masood-ul-Alam, Commun. Math. Phys. {\bf 162}(1994), 123; 
A. K. Masood-ul-Alam, Class. Quantum Grav. {\bf 5}(1988), 409; 
R. Beig and W. Simon, Commun. Math. Phys. {\bf 144}(1992), 373. 

\bibitem{LM}
L. Lindblom and A. K. M. Masood-ul-Alam, {\it Directions in 
General Relativity II}, ed. B. L. Hull and T. A. Jacobson
(Cambridge University Press, 1993). 

\bibitem{Simon}
W. Simon, Class Quantum Grav. {\bf 10},(1993)177; 
W. Simon, gr-qc/0204037; 
J. M. Heinzle, Class. Quantum Grav. {\bf 19}(2002),2835; 
J. M. Heinzle and C. Uggla, Class. Quantum Grav. {\bf 20}(2003),4567.

\bibitem{Chandra}
S. Chandrasekhar, {\it Introduction to the Study of Stellar Structure}
(University Chicago Press, 1939). 


\end{thebibliography}
\end{document}